\documentclass{article}

\usepackage{microtype}
\usepackage{graphicx}
\usepackage{subfigure}
\usepackage{textcomp}
\usepackage{siunitx}
\usepackage{booktabs} 

\usepackage{hyperref}

\usepackage[accepted]{icml2020}

\icmltitlerunning{TeleVital: Enhancing the quality of contactless health assessment}

\begin{document}

\twocolumn[
\icmltitle{TeleVital: Enhancing the quality of contactless health assessment}



\icmlsetsymbol{equal}{*}

\begin{icmlauthorlist}
\icmlauthor{Jithin Sunny}{man}
\icmlauthor{Joel Jogy}{man}
\icmlauthor{Rohan Rout}{man}
\icmlauthor{Rakshit Naidu}{man}
\end{icmlauthorlist}

\icmlaffiliation{man}{Manipal Institute of Technology, India}

\icmlkeywords{Machine Learning, Healthcare Systems, Contactless assessment, ICML}

\vskip 0.3in
]




\begin{abstract}
In the midst of rising positive cases of COVID-19, the hospitals face a newfound difficulty to prioritize on their patients and accommodate them. Moreover, crowding of patients at hospitals pose a threat to the healthcare workers and other patients at the hospital. With that in mind, a non-contact method of measuring the necessary vitals such as heart rate, respiratory rate and SPO$_2$ will prove highly beneficial for the hospitals to tackle this issue. This paper discusses our approach in achieving the non-contact measurement of vitals with the sole help of a webcam and further our design of an e-hospital platform for doctors and patients to attend appointments virtually. The platform also provides the doctor with an option to provide with voice-based prescriptions or digital prescriptions, to simplify the daily, exhausting routine of a doctor.
\end{abstract}

\section{Introduction}
\label{submission}
COVID-19 has taken the whole world by the storm. COVID-19(CoronaVirus Disease-19) is a disease caused by a new strain of the family of CoronaVirus. Most patients experience mild to moderate respiratory illness who recover without the need of any external treatment, although older individuals are likely to experience extreme illness, if they suffer from prevalent conditions like asthma or any other chronic respiratory disease. Testing patients for this disease has been a major issue as there is a dearth of testing kits. As there is no cure yet, the best way to slow down the transmission of the virus is to practice social distancing, although, this practice may have serious repercussions towards mental and physical health in the future.

Food shortage, unemployment crisis and the scarcity of healthcare professionals are just a few problems to name in these difficult times. Patients checking into hospitals regularly not only risk the infections to themselves, but also to the people around them. There is a need to reduce the burden on the healthcare systems around the world in order to match for the low number of healthcare workers. As the rate of the affected is growing exponentially, by the day, the hospital systems are now more overwhelmed than ever, which increases the liability of these machines. For the efficient management of this situation, it is necessary to prioritize who needs care and who does not. Today, it’s very difficult to do this because there are a lot of patients and not enough doctors. And this leads to overcrowding of hospitals and clinics and brings about an imbalance in the healthcare system. Therefore, it is necessary to handle this situation by analyzing the health of the patient at their homes itself \cite{VanGastel18, Iozzia17, Benari10, Mala17, Massaroni18}. This also prevents healthcare professionals from being directly exposed to the virus, which is a huge advantage in itself, because if the doctors and nurses get sick, then our first line of defense is lost.

A report\cite{anon20} from the Economic Times dated April 3rd stated that CRPF helpline gets 91 percent calls regarding coronavirus, lockdown information and relatively fewer  for rations and  healthcare. Public health organizations, hospitals and others on the frontlines of the COVID-19 response need to be able to respond to inquiries, cater to the public with up-to-date outbreak information, track exposure and quickly triage new cases. In a crisis like the COVID-19 pandemic, it’s not only important to deliver medical care but to also provide information to help people make decisions and prevent health systems from being overwhelmed. The healthcare workers are our first line of defence and we can't afford to lose them. Even with access to capable health care and social services, those afflicted with mental health conditions hesitate to avail the treatment due to the stigma surrounding the illness. Ultimately, the greatest lesson that COVID-19 can teach humanity is that we are all in this together and it is clear that a remote vital tracking system and online consultation and therapy is the need of the hour\cite{Benari10}.

Our product, TeleVital, is one step ahead in this pandemic situation by ensuring a clean, complete and contactless health assessment at the comfort of the patient's home. TeleVital is an AI-driven Chatbot tool designed to help people do preliminary analysis on various parameters right at their fingertips. This chatbot can also be accessed via premier social media platforms like Whatsapp and Facebook Messenger in their local languages to enhance the User Experience. 

We built a website that can help the patients to classify themselves as low risk, medium risk and high-risk category during these unprecedented times. If the patient is classified as a medium risk or a high-risk patient the application would suggest a line of therapy and medications based on Government of India’s Ministry of Health and Family Welfare(MoHFW) and Indian Council of Medical Research(ICMR) guidelines. An additional feature of a quick vitals analysis can be accessed through our platform. If the health of the patient deteriorates, the website is automated to alert the concerned authorities regarding the same.

As we target the issue of monitoring patients' vitals in a hospital, TeleVital performs the preliminary analysis on various parameters of the patient's health, prioritizing the user's comfort. Therefore, after consulting a few medical professionals, who are toiling hard to contain COVID-19 patients and after trying to understand the problem the doctors currently face, the experienced health workers suggested that the vitals play a major role in determining whether a patient needs hospitalization or not. The doctors recommended capturing vitals such as heart rate, respiratory rate, SpO$_2$ concentration in the blood and body temperature without any contact-based measurement tools such as a Pulse Oximeter which would potentially increase the spread of contact-based infections/diseases.

\section{Components}

The TeleVital platform consists of many features to facilitate both the patients as well as the healthcare professionals. 

\subsection{Chatbot}

The chatbot employed by the software uses a conditional approach to decision making. It is comprised of nodes containing information which is visible to the user and links them to other nodes. The links are decided either by choices made by the user while interacting with the chatbot or in cases where there are no choices, they are decided by a default case. This gives the option to the user to navigate through the chatbot as they wish to. They get an option to return to certain checkpoints while journeying through the chatbot in case of a change in mind. The chatbot is also an interactive shortcut for the user to explore different features that the app provides.

\subsection{Video Calling Platform}

Due to the spread of COVID-19 disease, it has been evident for needs to be able to take your regular visits to doctor from home and avoid visiting hospitals unless it is an emergency or advised by a doctor. For this very purpose, we built a platform for patients to have a 1-on-1 apppointments online and through video calling. Here, the patients are able to clearly explain their symptoms to the doctor,  and the doctor is able to use our TeleVital platform to measure the patient’s vitals and diagnose the patients in a remote manner. For the implementation of the video calling, we used WebRTC which enables real-time communication (voice, video) through a web browser. 

In order for WebRTC technologies to work, a request for your public-facing IP address is first made to a STUN server.The server then responds with your public-facing IP address which is communicated to the peer. These peers are also able to do the same thing using a STUN or TURN server and can tell you what address to contact them at as well. Signaling data “channels” are then dynamically created to detect peers and support peer-to-peer negotiations and session establishment. Once two or more peers are connected to the same “channel”, the peers are able to communicate and negotiate session information. This process is similar to the publish/subscribe pattern. Once the answer is received, a process occurs to determine and negotiate the best of the ICE candidates gathered by each peer. Once the optimal ICE candidates are chosen, essentially all of the required metadata, network routing (IP address and port), and media information used to communicate for each peer is agreed upon. The network socket session between the peers is then fully established and active. Next, local data streams and data channel endpoints are created by each peer, and multimedia data is finally transmitted both ways using whatever bidirectional communication technology is employed.

\subsection{Vitals Measurement}

For the measurement of vitals, the frames are captured via the web camera and these frames are sent to a real-time database hosted on Firebase in base64 format, which are later converted back to PIL image format and manipulated accordingly to calculate Heart Rate, Respiratory Rate and SpO$_2$ levels in the blood. 

\textbf{Heart Rate(HR)}

Heart rate is one of the most important vitals for studying a person’s physiological state and can help in realizing a person’s aerobic capacity or their risk in getting a heart attack. Apart from these, one’s heart rate plays a key role in estimating early symptoms of COVID-19, it has been proved. In standard conventions such as at hospitals or at home, an electrocardiogram (ECG) or a pulse oximeter can be used to measure one’s heart rate but in cases of non-contact measurement, it is also possible to have it measured using a webcam or a mobile phone camera. In our approach, we were able to measure heart rate with an error rate of 3.1 $\pm$ 0.3 bpm in still video and 2.2 $\pm$ 1.8 bpm in video with movement\cite{isabelbush}.

Heart Rate is defined as the number of a person’s heart beats per minute (bpm). A normal heart rate of an adult ranges from 60 to 100 bpm. To determine the heart rate, image frames containing the person’s face are read. A monitoring algorithm for face detection is applied to select the Region of Interest (ROI). Classifiers are used on the frame to detect the coordinates of boundary for ROI selection and hence the region of interest is acquired. An RGB signal extraction is carried out on the frame for accurate computations. Then the ROI is put under an algorithm of signal detrending to remove drifts and noise. On the resulting frame, some filtering and normalisation are carried out along with matrix calculations. These calculations result in heart rate in beats per minute (bpm).

\textbf{Respiratory Rate(RR)}

The number of breaths taken by a person in a minute is termed as the Respiratory Rate for the person. 
A normal respiratory rate of an adult ranges from 12 to 20 breaths per minute. For the calculation of respiratory rate, the oscillations of the face while inhaling and exhaling are studied for relative displacement. The first step in this process is capturing frames of the face of the person to be tested. The frames are sampled to get a clear picture of the change in pixel intensities. Then the frames are preprocessed to find the required template which has the least noise and sharper features. The region of interest (ROI) is determined from the template. Then basic signal processing techniques are applied on to the ROI to extract the frequency of the change in intensity of pixel. The most dominant frequency is selected as the respiratory rate and is displayed in frequency per minute. 

Traditional methods record the maximum change in frequency. Our novel approach keeps track of both, the mean and the maximum change of frequency over a certain number of frames. These values are then reduced to a single value by calculating the mean. The results have suggested our approach to be more accurate than the previous approaches.

\textbf{SpO$_2$ Levels}

Peripheral capillary oxygen saturation (SpO$_2$), which is an estimate of the amount of oxygen in the blood is one of the key vitals to be monitored. For SpO$_2$ calculation, the behaviour of components of white light is monitored when a finger is placed between a source of white light and the camera. The captured frame is preprocessed by splitting it into red and blue color channels. Then specific numerical computations involving the mean and standard deviations are carried out on these channels. These give out the concentration of oxygen in blood in percentage. 

Firstly, the image is resized to 320x240 and split into its red and blue channels. The Mean of red and blue components and the Standard deviations of red and blue components are calculated for each image\cite{Kanva14}. The mean values correspond to the DC components and standard deviations correspond to the AC components. 

Then SpO$_2$ is calculated as \[ SpO_{2} = A - (B \times ((S_{red} /M_{red})/(S_{blue}/M_{blue}))) \]
where S$_x$ stands for Standard Deviation of channel $x$ and M$_x$ stands for Mean of channel $x$.

Normal SpO$_2$ levels of an adult ranges from 95 to 100 percent.

\subsection{Digital Prescription}

To ease doctors with providing patients hand-written prescriptions and also evading the normal contact-based prescription methods, the digital prescription portal enables the doctors to voice record prescription for patients. When the doctor visits the website, they can record their voice by clicking on the record button and record all necessary details like name, age, gender of the patient, their symptoms, the diagnosis, the prescription and also suggest professional medical advice. This audio file is converted into text, which is filtered and displayed in text boxes namely Age, Name, Sex, Symptoms, Diagnosis, Prescription, and Advice. If the doctor needs to edit or add something, they can do it here manually and then finally they can generate the pdf of the report by clicking on the generate PDF button which can be viewed, downloaded or mailed to the patients mail ID.

\subsection{Patient Prioritizer}

Understanding the rapid increase in the cases of COVID-19, the Patient Prioritizer page enables hospitals in identifying the high risk patients and triaging the patients accordingly. The comparison between the patients is done based on 14 parameters namely: Age, Gender, Height, Weight, Heart rate, Respiratory rate, SpO$_2$ levels, Body temperature, Cough, Sore throat, Breathing difficulty, Tiredness/Fatigue, Pre-existing medical conditions like asthma and Pregnancy. Weights are assigned to every parameter(adhering to the WHO guidelines) which are used to calculate the final score of the patients. The doctor can view the patients' name and their corresponding scores. The list is arranged in a descending order so that the professionals can also search for the patient by their name. The patient with the highest score requires the maximum amount of medical attention. The doctor can edit the already registered patients details and searching the patients by name and age would retrieve all of the patient's previous records. These records displayed on the screen can be modified or deleted by the doctor accordingly.

\section{Experiments}

We tabulated readings using both, a Pulse Oximeter(Table 1) and our TeleVital portal(Table 2) to compare the efficiency of our product. This experiment was conducted on three volunteers and the following parameters were recorded: Heart Rate(HR), Respiratory Rate(RR) and SpO$_2$ concentration.

\begin{table}[t]
\caption{Pulse Oximeter Readings}
\label{pulseox-table}
\vskip 0.15in
\begin{center}
\begin{small}
\begin{sc}
\begin{tabular}{lcccr}
\toprule
HR & RR & SpO$_2$\\
\midrule
Person 1: \\
92 & 15 & 96 \\
95 & 19.5 & 97 \\
90 & 15.3 & 96 \\
Person 2: \\
100 & 19 & 92 \\
86 & 18 & 93 \\
81 & 12 & 97 \\
Person 3: \\
82 & 12 & 90 \\
86 & 17.4 & 95 \\
84 & 14.6 & 94 \\
\bottomrule
\end{tabular}
\end{sc}
\end{small}
\end{center}
\vskip -0.1in
\end{table}

\begin{table}[t]
\caption{TeleVital readings}
\label{televital-table}
\vskip 0.15in
\begin{center}
\begin{small}
\begin{sc}
\begin{tabular}{lcccr}
\toprule
HR & RR & SpO$_2$\\
\midrule
Person 1: \\
83 & 17.78 & 95.24 \\
86 & 17.89 & 95.1 \\
85 & 17.53 & 94.97 \\
Person 2: \\
91 & 19.09 & 94.92 \\
86 & 18.5 & 94.12 \\
85 & 16.9 & 95.06 \\
Person 3: \\
78 & 14.3 & 94.05 \\
85 & 17.48 & 96.47 \\
79 & 19.23 & 93.56 \\
\bottomrule
\end{tabular}
\end{sc}
\end{small}
\end{center}
\vskip -0.1in
\end{table}

\section{Conclusions}

Through our webcam-based vital measurement approach, we have clearly proved that a person can take the vital test from the comfort of their homes and understand if they show symptoms of COVID-19. This will enable the hospitals to focus and prioritise on the patients showing critical vitals. While the accuracy of the results depend directly on quality of the camera feed, we have been able to understand from the webcams we have tried for various laptop brands, the accuracy still proved to be comparable to the reading from the respective instruments for measuring the vitals. We also found that as bounding box noise is increased, choosing an ROI by segmenting out facial pixels helped to diminish the noise effects and keep the outliers low.

Future studies could involve taking video in low-light and with a cluttered background or with occlusions or multiple subjects in the frame to quantify how much bounding box noise is realistically present in such situations. This could help to determine if segmentation is necessary. 

A side feature that we are working on is a Dialogflow chatbot\cite{Shah18} that revolutionizes the contactless health measurement with a  voice-based response to assist the needy and the disabled.

Another aspect is enhancing the User Interface(UI) and User Experience(UX) of the TeleVital platform. Given that a majority of the users will be patients or doctors, improving the UI could help in enriching the moods of the users.

We also plan to integrate facilities for measurement of different parameters like height and weight [through Augmented Reality] so that the reach of this application goes beyond the COVID pandemic.

Another future scope of this project could be to develop this platform as a kiosk to be used at office at the entry stops being employees enter their offices. The cameras on the kiosk could first identify if the employee is wearing a fask mask, then proceed to take a vital test and finally take their attendance using a face recognition system before they are let into the offices. These kiosks or hardware robots could be stationed at public places also, like airports or malls to perform contactless health assessments. Source code for implementations of the algorithms discussed in this paper may be found at \url{https://github.com/jithinsunnyofficial/TeleVital-MicrosoftHack}

\bibliography{televital}


\bibliographystyle{icml2020}

\end{document}